\documentclass[conference]{IEEEtran}
\IEEEoverridecommandlockouts
\usepackage{graphicx}
\usepackage{xcolor}
\usepackage{acronym}
\usepackage{multirow}
\usepackage[space]{cite}
\usepackage{algorithm}
\usepackage[noend]{algpseudocode}
\usepackage{multirow}
\usepackage{booktabs}
\usepackage{mathtools}
\usepackage{amsmath}
\usepackage{cleveref}
\usepackage{amsmath}
\usepackage{xfrac}
\usepackage{subfigure}
\usepackage{arydshln}

\begin{document}

\title{LOGAN: High-Performance GPU-Based $X$-Drop Long-Read Alignment}

\acrodef{NW}[NW]{Needleman--Wunsch}
\acrodef{SW}[SW]{Smith--Waterman}
\acrodef{GPU}[GPU]{Graphical Processing Unit}
\acrodef{GCUPS}[GCUPS]{Giga Cell Updates Per Second}
\acrodef{GIPS}[GIPS]{Giga Instructions Per Second}
\acrodef{SM}[SM]{Streaming Multiprocessor}
\acrodef{FPGA}[FPGA]{Field Programmable Gate Array}
\acrodef{HBM}[HBM]{High Bandwidth Memory}

\author{\IEEEauthorblockN{
Alberto Zeni\IEEEauthorrefmark{1},
Giulia Guidi\IEEEauthorrefmark{2}\IEEEauthorrefmark{3}, 
Marquita Ellis\IEEEauthorrefmark{2}\IEEEauthorrefmark{3}, 
Nan Ding\IEEEauthorrefmark{3},
Marco D. Santambrogio\IEEEauthorrefmark{1}, \\
Steven Hofmeyr\IEEEauthorrefmark{3},
Ayd{\i}n Bulu\c{c}\IEEEauthorrefmark{2}\IEEEauthorrefmark{3}, 
Leonid Oliker\IEEEauthorrefmark{3},
Katherine Yelick\IEEEauthorrefmark{2}\IEEEauthorrefmark{3}}\vspace{.25cm}

\IEEEauthorblockA{
\IEEEauthorrefmark{1}Dipartimento di Elettronica, Informazione e Bioingegneria, Politecnico di Milano, Milan, Italy \\
\IEEEauthorrefmark{2}Department of Electrical Engineering and Computer Science, University of California at Berkeley, Berkeley, CA, USA \\
\IEEEauthorrefmark{3}Computational Research Division, Lawrence Berkeley National Laboratory, Berkeley, CA, USA}
\textbf{Availability}: https://github.com/albertozeni/LOGAN\\
\textbf{Contact}: \hspace{.035cm}alberto.zeni@mail.polimi.it, gguidi@lbl.gov}

\maketitle
\thispagestyle{plain}
\pagestyle{plain}

\begin{abstract}
Pairwise sequence alignment is one of the most computationally intensive kernels in genomic data analysis, accounting for more than 90\% of the runtime for key bioinformatics applications.
This method is particularly expensive for \textit{third-generation} sequences due to the high computational cost of analyzing sequences of length between 1Kb and 1Mb.
Given the quadratic overhead of exact pairwise algorithms for long alignments, the community primarily relies on approximate algorithms that search only for high-quality alignments and stop early when one is not found.
In this work, we present the first GPU optimization of the popular $X$-drop alignment algorithm, that we named LOGAN. 
Results show that our high-performance multi-GPU implementation achieves up to 181.6 GCUPS and speed-ups up to 6.6$\times$ and 30.7$\times$ using 1 and 6 NVIDIA Tesla V100, respectively, over the state-of-the-art software  running on two IBM Power9 processors using 168 CPU threads, with equivalent accuracy. 
We also demonstrate a 2.3$\times$ LOGAN speed-up versus ksw2, a state-of-art vectorized algorithm for sequence alignment implemented in minimap2, a long-read mapping software. 
To highlight the impact of our work on a real-world application, we couple LOGAN with a many-to-many long-read alignment software called BELLA, and demonstrate that our implementation improves the overall BELLA runtime by up to 10.6$\times$.
Finally, we adapt the Roofline model for LOGAN and demonstrate that our implementation is near optimal on the NVIDIA Tesla V100s.

\end{abstract}

\section{Introduction}
\label{abs:intro}

Pairwise alignment is one of the most commonly used workhorses of sequence analysis.  
It is used to correct raw sequencer reads, assemble them into more complete genomes, search databases for similar sequences, and many other problems. 
The optimal solutions for this problem require quadratic time (i.e. they take $O(mn)$ time for aligning a sequence A of length $m$ and a sequence B of length $n$). 
Namely, Needleman--Wunsch (NW)~\cite{needleman1970general} is used to find the best global alignment by forcing the alignment to extend to the endpoints of both sequences. 
Alternatively, Smith--Waterman (SW)\cite{smith1981identification} computes the best local alignment by finding the highest scoring alignment between continuous subsequences of the input sequences.

The popular $X$-drop~\cite{zhang2000greedy} algorithm avoids the full quadratic cost by searching only for high-quality alignments, and can be viewed as an approach to accelerate both NW and SW. Most applications of alignment will throw out low quality alignments, which arise when the two strings are not similar. Instead of exploring the whole $m \times n$ space, the $X$-drop algorithm searches only for alignments that results in a limited number of edits between the two sequences. $X$-drop keeps a running maximum score and does not explore  cell neighborhoods whose score decreases by a user-specified parameter $X$. It gets its performance benefits from searching a limited space of solutions and stopping early when a good alignment is not possible. 

Zhang et al.~\cite{zhang2000greedy} proved that, for certain scoring matrices, the $X$-drop algorithm is guaranteed to find the optimal alignment between relatively similar sequences. In practice, the algorithm eliminates searches between sequences that are clearly diverging. This feature is especially effective for many-to-many alignment problems when there is an attempt to align many sequences to many other possibly matching sequences, i.e., the cost is high as is the possibility that some pairs will not align. With $X$-drop, any spurious candidate pair is readily  eliminated because the optimal score quickly drops. Consequently, $X$-drop and its variants are the algorithm of choice in some of the most popular sequence mapping software including BLAST~\cite{altschul1990basic}, LAST~\cite{kielbasa2011adaptive}, BLASTZ~\cite{schwartz2003human} with $Y$-drop, and minimap2~\cite{li2018minimap2} with $Z$-drop. 

Although $X$-drop is a heuristic for cutting the cost of alignment, it also produces good quality results, which are sometimes better than a more complete search. Frith et al.~\cite{frith2010parameters} show that  a large $X$ does not necessarily produce better alignments.
Without the $X$-drop feature, the alignment algorithm can incorrectly glue two independent local alignments into a large one. 
For example, consider two sequences, one of the form $\textrm{S = A-B-C}$ and other of the form $\textrm{R = A-D-C}$. Since the regions A and C produce high-scoring alignments, likely a high $X$ would incorrectly determine that  $\textit{score}(\textrm{S,R}) > \textit{max}(\textit{score}(\textrm{A,A}), \textit{score}(\textrm{C,C}))$ provided that $\textrm{B}$ and $\textrm{D}$ regions are short enough.

Although there are numerous GPU implementations of the full $O(mn)$ SW and NW algorithms that often achieve impressive computational rates (measured in CUPS or cell updates per second), they are rarely incorporated into high-impact genomics pipelines due to their quadratic complexity. By contrast, a GPU implementation of $X$-drop is notably  missing from the literature despite its benefits and popularity.
This is likely due to the increased complexity of implementing $X$-drop efficiently on a GPU, compared with NW and SW methods, because of the dynamic nature of the computation, its adaptive band, and the need to check for completion.

Our main contributions are:
\begin{itemize}
    \item We present the first high-performance, multi-GPU implementation of the $X$-drop algorithm, named LOGAN,
    which achieves significant speed-ups over leading versions on state-of-the-art processors.
    \item We integrate LOGAN within BELLA, a long-read many-to-many overlapping and alignment software and demonstrate performance improvements up to $10\times$.
    \item We adapt the Roofline Model to LOGAN implementation and underlying hardware, and demonstrate that performance is near optimal on NVIDIA Tesla V100s.
\end{itemize}

A key aspect of our implementation is combining different levels of parallelism.
Specifically, we implement the {\em intra-sequence} parallelism via dynamic thread scheduling and in-warp parallelism, while accomplishing {\em inter-sequence} parallelism by assigning each GPU block to a single alignment.
Finally, we carry out parallelism across multiple GPUs through a load balancer.\\

The remainder of the paper is organized as follows.
\Cref{abs:rel} provides an overview of the related work, while \Cref{abs:backgr} describes the original software algorithm we port on GPU.
\Cref{abs:impl} describes our implementation and optimizations.
\Cref{abs:ki} presents LOGAN integration within BELLA~\cite{guidi2018bella}, a long-read overlapping and alignment software.
\Cref{abs:res} illustrates our experimental results, while \Cref{abs:roof} describes the Roolfline model used to analyze our implementation. 
Finally, \Cref{abs:concl} summarizes our contributions and outlines  future work.

\section{Related Work}
\label{abs:rel}

The majority of hardware acceleration efforts for pairwise alignment have focused on the \ac{SW} and \ac{NW} algorithms. These find exact alignments and have quadratic complexity in the lengths of the reads. Along with some of the most successful NW and SW acceleration efforts, we review the few efforts to accelerate heuristics more similar to our own. Though they are more generally applicable, exploiting GPU parallelism in these heuristics is more challenging due to their adaptive nature. As a common success metric, we report \ac{GCUPS}, as reported by the original work, throughout this section. 
It is important to keep in mind that the GCUPS rates presented in this section were collected by the respective authors using different architectures than examined in our study. 
In Section~\ref{abs:res}, we collect comparative performance data with the ksw2 algorithm on equivalent platforms.

The implementation of Michael Farrar~\cite{farrar2006striped}, which is adopted in Bowtie2~\cite{langmead2012fast}, stands out amongst software implementations of the SW algorithm.
It leverages SIMD instructions and reaches performance of more than 20 GCUPS on an Intel Xeon Gold with 40 CPU threads. 
The same software implementation has been optimized for the PlayStation 3 processor and the IBM QS20 architecture~\cite{szalkowski2008swps3}, achieving performance of $15.5$ and $11.6$ GCUPS, respectively.

CUDASW++3~\cite{liu2013cudasw++} accelerates the SW algorithm combining SIMD instructions and GPU parallelism.
The implementation achieves up to $185.6$ GCUPS when aligning reads with length less than $400$ characters.
However, the performance drops significantly when the sequence length exceeds $400$ characters. Additionally, when running only using the GPU, their maximum attained performance is $68$ GCUPS (roughly $\sfrac{1}{3}$ of their peak performance). 

Muhammadzadeh presented MR-CUDASW++ \cite{mr-cudasw2014}, which was inspired by CUDASW++3 but optimized for ``medium length'' reads. Muhammadzadeh compares MR-CUDASW++ to other tools, with CUDASW++3 as its closest contender, across sequences lengths of $1$K, $10$K, and $100$K. MR-CUDASW++ achieved speedups of $1-2\times$ over CUDASW++3. The results were below 85 GCUPS using an NVIDIA Tesla V100, which is the same GPU used for our benchmarks.
Li et al.~\cite{li2007160} accelerate the SW algorithm achieving a speed-up of over $160\times$ compared to software implementation using an Altera Nios II \ac{FPGA}.
Nevertheless, the performance of their proposed solution is comparable to existing optimized software implementations.
The SW implementation of Di Tucci et al.~\cite{di2017architectural}, running on a Xilinx Virtex 7 and a Kintex Ultrascale platforms, achieves up to $42.5$ GCUPS and a speed-up of $1.7\times$ over the state-of-the-art FPGA implementation.
However, this work is limited to aligning short sequences that have a number of characters not exceeding the number of processing elements in the architecture.

A recent work by Turakhia et al.~\cite{turakhia2018darwin}, Darwin, exploits FPGAs to speed up the alignment process achieving up to $45$ GCUPS.
Darwin uses a seed-and-extend heuristic (GATC) that performs the extension stage in the seed-and-extend paradigm in which Dynamic Programming (DP) is used around the seed hit to obtain local alignments similar to SW.

Feng et al.~\cite{FengIcpp2019} recently presented accelerator-based optimizations for minimap2~\cite{Minimap2}. 
Leveraging the GPU architecture, they accelerate minimap2's {\it seed-chain-extend} pairwise alignment algorithm, which is quadratic in the length of the reads when computing traceback and linear otherwise. 
They reported performance of $96.5$ GCUPS (a $7.1\times$ speed-up over the minimap2's SIMD software implementation).
Our $X$-drop alignment algorithm in this study computes a similar heuristic to minimap2's pairwise alignment. 
In~\Cref{abs:res}, we compare our work with ksw2 \cite{suzuki2018introducing} (i.e., minimap2's alignment kernel), 
showing that LOGAN achieves higher performance in terms of GCUPS than both ksw2 and the performance reported by Feng et al. for their GPU-accelerated implementation. 

Despite a large number of sophisticated implementations, the overwhelming majority of the proposed hardware accelerations studies implement the exact SW or NW algorithms.
We believe the although $X$-drop algorithm is the most practical choice for targeting large-scale alignments, it requires more challenging parallelization than the original SW or NW methods. Though relatively unexplored due to this challenge, we demonstrate that GPU  optimization results in significant acceleration. 

\section{Background}
\label{abs:backgr}

This section provides an overview of the $X$-drop implementation proposed by Zhang et al.~\cite{zhang2000greedy} and implemented in the SeqAn library~\cite{doring2008seqan}, a C++ library for sequence analysis. 
First, we review the formal definition of alignment. 
A {\it pairwise alignment} of sequences {\it s} and {\it t} over an alphabet $\Sigma$ is defined as the pair $(s',t')$ such that $s',t' \in \Sigma \cup \{-\}$ and the following properties hold:
\begin{enumerate}
  \item  $|s'| = |t'|$ = $l$ 
  \item  $\forall_{i=1}^{l} \ \  s_i' \neq -\ \ OR\ \ t_i'  \neq -$ 
  \item  Deleting all ``$-$'' from $s'$ yields $s$, and deleting all ``$-$'' from $t'$ yields $t$.  
\end{enumerate}
A {\it scoring scheme} is used to distinguish high-quality alignments from the many (valid) alignments of a given pair of sequences. Scoring schemes generally reward matches and penalize mismatches, insertions, and deletions.

\subsection{The $X$-drop Algorithm}

\begin{figure}[t]
\centering
\vspace{-.2cm}
\includegraphics[width=\columnwidth]{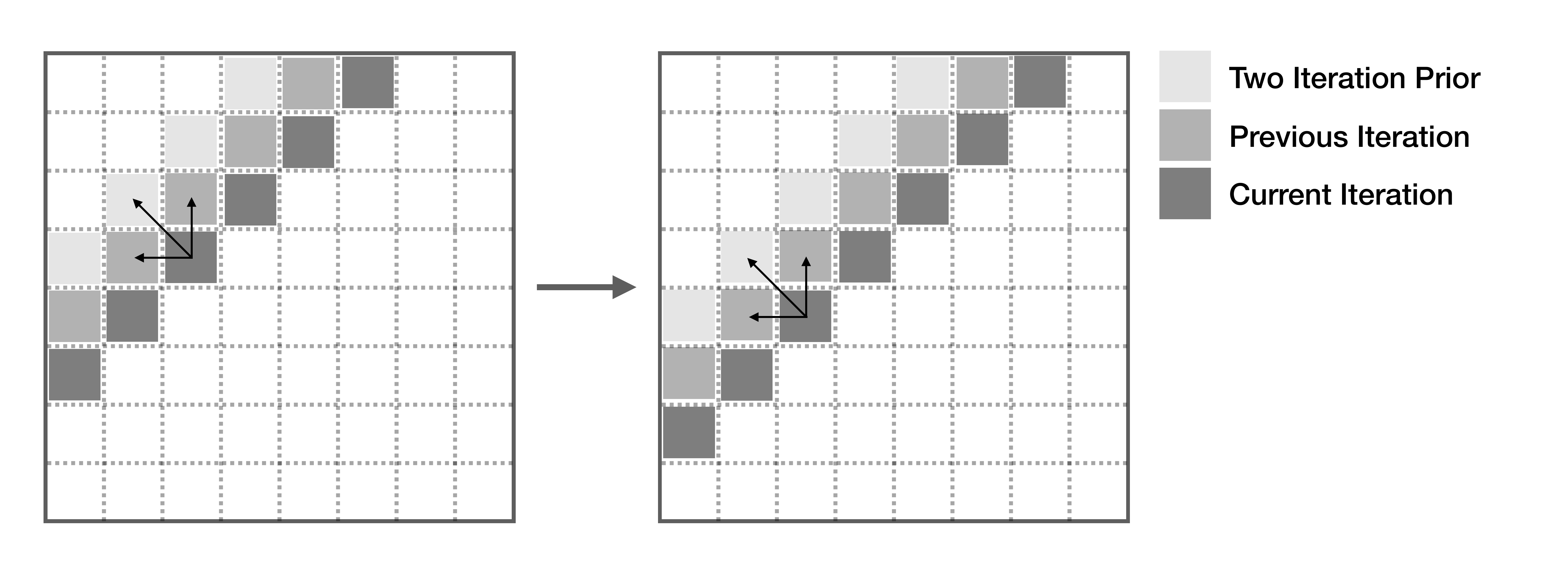}
\caption{
    Each cell at the current iteration has two dependencies on cells from the previous iteration and one dependency on a cell at two iteration prior.
}
\label{fig:dependencies}
\end{figure}

\begin{figure}[t]
\centering
\vspace{-.15cm}
\includegraphics[width=\columnwidth]{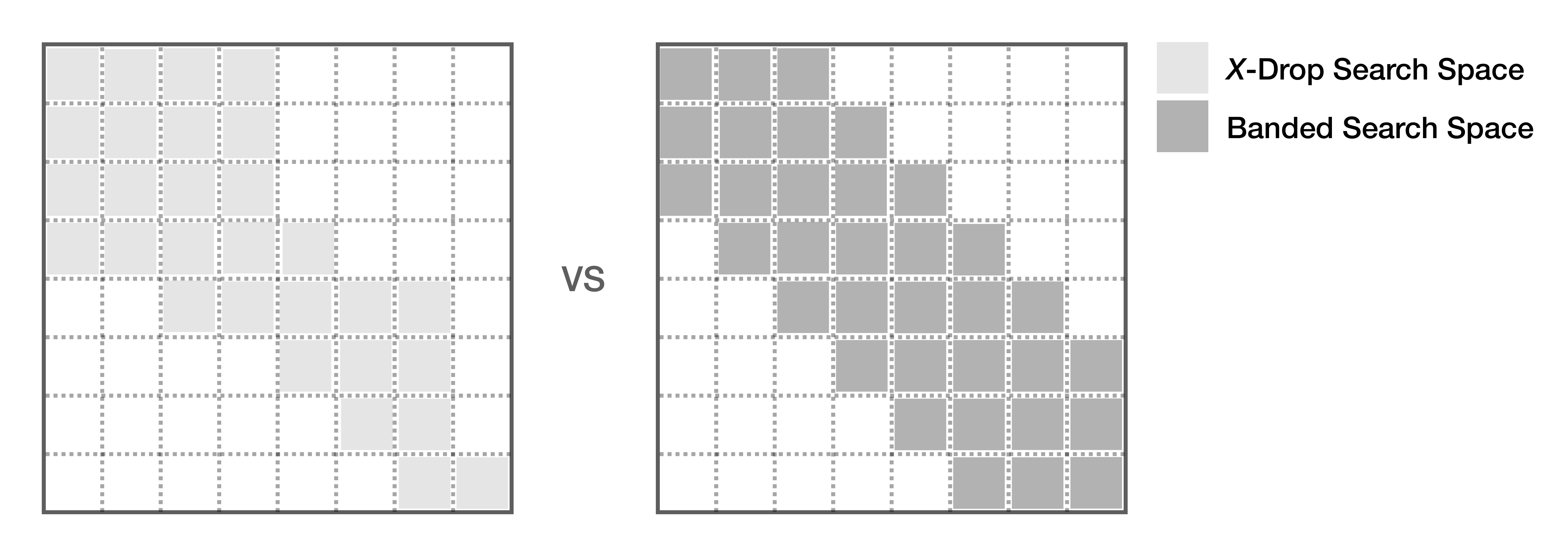}
\caption{
    Comparison between the search space of an $X$-drop alignment algorithm versus the search space of a banded-alignment algorithm.
}
\label{fig:searchspace}
\end{figure}

Given two DNA sequences $A = a_1a_2 \dots a_m $ and  $B = b_1b_2  \dots b_n$ of length $m$ and $n$, the goal of the $X$-drop algorithm is to find the highest-scoring {\em semi-global} alignment between $A$ and $B$ of the forms $a_1a_2 \dots a_i$ and $b_1,b_2 \dots b_j$, for some $i \leq m$ and $j \leq n$ that are chosen to maximize the score. 


For a given $i$ and $j$, we define $S$ as the alignment matrix and $S(i,j)$ as the alignment score between $A$ and $B$.
A positive {\em match} score is added to $S(i,j)$  for each pair of identical nucleotides. If nucleotides do not match, the algorithm can either subtract a {\em mismatch} score to $S(i,j)$  and move diagonally or subtract a {\em gap} score and move horizontally (gap into the vertical sequence) or vertically (gap into the horizontal sequence) in the dynamic programming grid. 
More formally, each cell of the alignment matrix $S$ is computed as follows:

\vspace{.35cm}
\resizebox{.925\columnwidth}{!}{%
$S(i,j) = 
\left
\{
\begin{matrix*}[l]
S(i-1,j-1) + match  & $if $i >0,j>0$ and $ a_{i}=b_{j}\\ 
S(i-1,j-1) + mismatch & $if $i >0,j>0$ and $ a_{i}\neq b_{j}\\ 
S(i,j-1) + gap & $if $j>0\\ 
S(i-1,j) + gap & $if $i>0
\end{matrix*}
\right.$}
\vspace{.35cm}

Figure~\ref{fig:dependencies} shows the three dependencies of a cell during the computation: two dependencies on cells from the previous iteration and one dependency on a cell at two iterations prior. Note that \ac{SW}, \ac{NW}, as well as the majority of their heuristic implementations show these dependencies.
SW and NW compute the entire $S$ matrix to find the optimal alignment. 
This quadratic algorithm is extremely inefficient in the case of either misalignment or when aligning almost identical sequences.
A misalignment could happen, for example, when two sequences have a tiny region in common due to a genomic repetition. 
The SW algorithm would spend significant computational resources calculating  the entire dynamic programming (DP) matrix and report a very poor alignment score between the two sequences.  
On the other hand, SW or NW on two almost identical sequences would compute the whole DP matrix with no additional benefit, since the optimal alignment score would remain close to the  diagonal of the DP matrix.

The concept of the $X$-drop termination consists of halting the computation if the alignment score drops more than $X$ below the best alignment score $\sigma$ seen so far for that pair of sequences.
$\sigma$ is potentially updated at each anti-diagonal iteration.
If $S(i,j) < \sigma - X$, we set the cell $S(i,j)$ equal to $-\infty$ and no longer consider that cell for the subsequent iterations of $S$.
The cells set to $-\infty$ are used to compute the lower and upper bound for the next anti-diagonal iteration. 
This approach limits the anti-diagonal width, reducing the search space of the algorithm, and automatically provides a termination condition. 
The $X$-drop algorithm is particularly efficient when two sequences do not align.
A pseudo-code of the $X$-drop algorithm is shown in Algorithm~\ref{alg:single_al}.

\begin{algorithm}[t]
\caption{Pairwise alignment of $S_q$ and $S_t$ with $X$-drop}\label{alg:single_al}
\begin{algorithmic}[1]
\Procedure{PairwiseAlignment}{$S_q$,~$S_t$,~$X$}
    \State $\textit{A1,~A2,~A3}$ \Comment{Create anti-diagonal}
    \State $\textit{best} \gets \textit{0}$ \Comment{Initialize best score to 0}
    \While{{$A1.size()\neq 0$}}  \Comment{DP matrix}
    \State $A1 \gets A3$.    \Comment{Anti-diagonal swap}
    \State $A2 \gets A1$.
    \State $A3 \gets A2$.
    \State ComputeAntiDiag($A1,~A2,~A3$)
    \State $best \gets A1.max()$
    \For{$k \gets 0$ to $A1.size()$} 
    \If {$A1[k] = -\infty$}
    \State ReduceAntiDiagFromStart($A1$)
    \EndIf
    \EndFor
    \For{$k \gets A1.size()$ to $0$} 
    \If {$A1[k] = -\infty$}
    \State ReduceAntiDiagFromEnd($A1$)
    \EndIf
    \EndFor
    
    \EndWhile
    \State \textbf{return}($best$) \Comment{$X$-drop termination}
\EndProcedure
\end{algorithmic}
\end{algorithm}

Note that $X$-drop should not be confused with the popular banded-SW method.
This approach constrains the search space to a fixed band along the diagonal, regardless of the drop in the score.
The areas of the $m \times n$ dynamic programming grid explored by these algorithms are characteristically different from $X$-drop's search space, which is reminiscent of a rugged band with  changes in the length of each anti-diagonal, as shown in Figure~\ref{fig:searchspace}.
To better understand the difference in practice, consider two sequences that have very high (over 50$\%$) differences in terms of substitutions but have no indel (insertion or deletion) differences. 
The optimal path would be along the diagonal because both a mismatch and match move the cursor in both sequences. 
$X$-drop will correctly terminate the search early due to a significant drop in the score whereas banded-SW would explore the entire band. 


\section{Implementation}
\label{abs:impl}

The key aspect of our implementation is exploiting as many levels of parallelism as possible on the \ac{GPU}.
Intra-level parallelism is achieved by dynamically scheduling the threads based on the value of $X$ and through the use of in-warp parallelization to find the maximum of the anti-diagonals.
To achieve inter-level parallelism, we implement the parallel execution of multiple alignments by assigning each GPU block to a single alignment.
Multi-GPU parallelism is obtained by implementing a GPU load balancer that adapts the execution of LOGAN to leverage multiple \acp{GPU}.

In this section, we describe the design of our implementation.  \Cref{abs:ki} then presents our optimized kernel integration within BELLA, a real-world application.
Note that we refer to GPU threads and GPU blocks simply as {\em threads} and {\em blocks}.



\subsection{Intra-Sequence Parallelism}
\label{sec:intra}

We first consider the intra-sequences parallelism, which is the parallelization of a single pairwise alignment and its anti-diagonals computation.
Given that the $X$-drop algorithm we decided to port to GPU does not perform alignment trace-back, we do not store the entire alignment matrix on the device for each alignment.
Therefore, we can reduce the memory footprint of our kernel on the GPU by storing only three anti-diagonals per alignment: current, previous, and two iterations prior, as highlighted in Figure~\ref{fig:dependencies}.

\begin{figure}[t]
\centering
\vspace{-.35cm}
\includegraphics[width=.5\columnwidth]{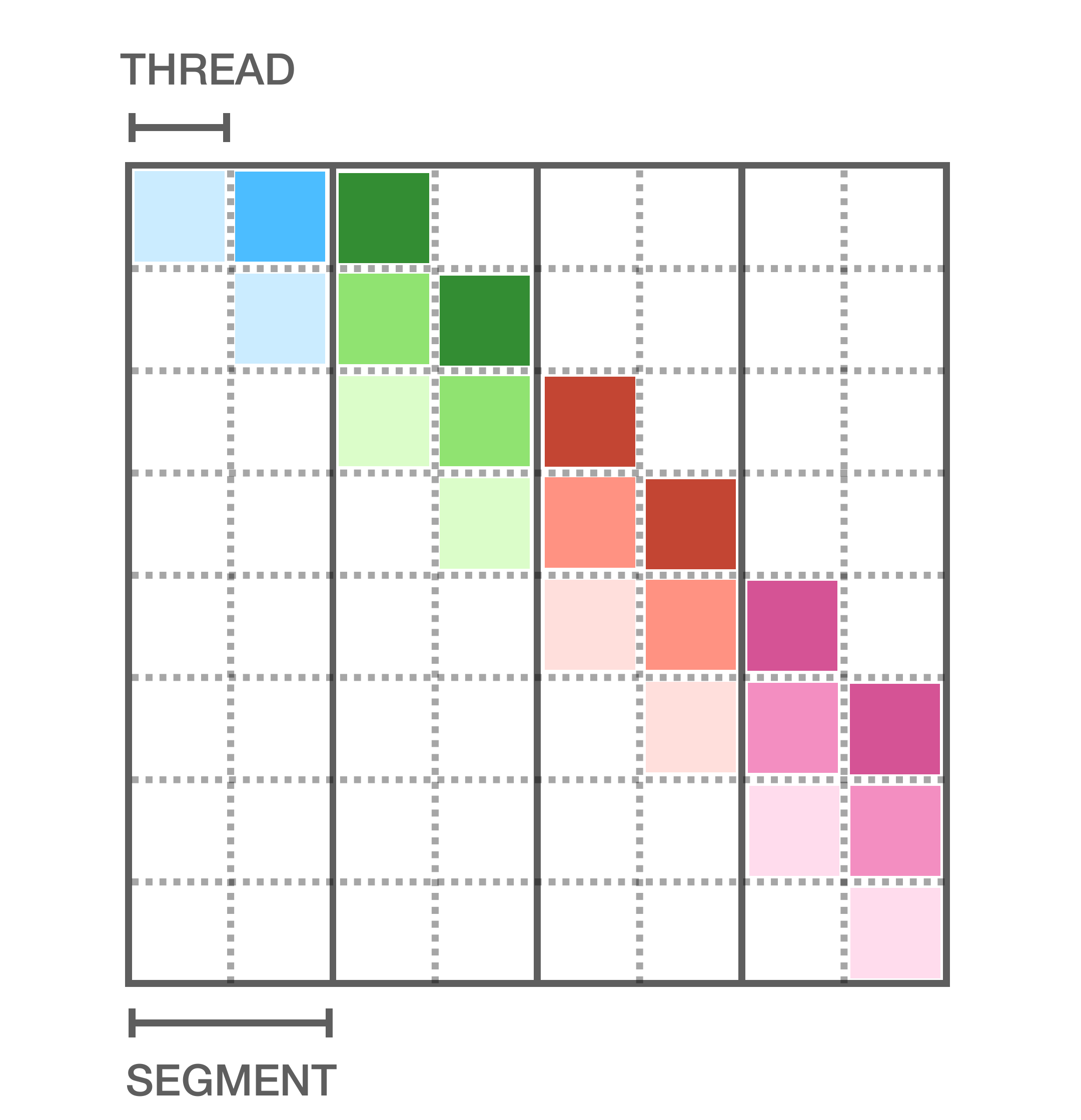}
\caption{
    Each anti-diagonal is divided in segments, whose width is equal to the number of $threads$ scheduled in a block
}
\label{fig:antidiag}
\end{figure}

Similarly to SW and NW algorithms, we compute the cell updates of each anti-diagonal of the alignment matrix in parallel. 
Each anti-diagonal cell has three dependencies on cells from anti-diagonals at previous iterations.
Nevertheless, we can leverage the independence between cells belonging to the same anti-diagonal.
To compute an anti-diagonal update in parallel, we assign each cell to a GPU thread and compute them independently, where a GPU block can schedule up to 1024 threads.
To overcome this limitation and ensure the computation of any anti-diagonal length, we split each anti-diagonal into {\em segments}, as shown in Figure~\ref{fig:antidiag}.
The anti-diagonal is split into segments whose width is equal to the number of threads within a block. 
Once a segment of the anti-diagonal is completed, the kernel initiates the computation of the subsequent segment.
This process is repeated until the entire anti-diagonal is computed.


For each cell, the corresponding thread sets the score to $-\infty$ if the cell score drops $X$ below the global maximum of the alignment matrix, that is {\em best} in Algorithm~\ref{alg:single_al}.
The overall maximum score is a shared variable within the considered block.
The global maximum of the scoring matrix is updated after any anti-diagonal has been completely calculated since it needs to consider the newly computed scores.
Computing the global maximum naively would significantly slow down the execution since it would require serial comparison of each cell with the all others in a given anti-diagonal.
Thus we speed up this process by computing the maximum anti-diagonal score via a parallel reduction. 

\begin{algorithm}[t]
\caption{Computation of the anti-diagonal in parallel}\label{alg:antidiag_al}
\begin{algorithmic}[1]
\Procedure{AntiDiag}{$A1,A2,A3,S_q,S_v,X,best$}
\State $tid \gets \mathit{threadID}$
\While {$tid < A1.size()$}
\If {$S_q[tid] == S_v[tid]$}
\State $A1[tid] \gets A3[tid-1] + \textbf{\em match}$
\Else
\State $A1[tid] \gets A3[tid-1] + \textbf{\em mismatch}$
\EndIf
\State $tmp \gets \textrm{\textbf{max}}(A2[tid] + \textbf{\em gap},~A2[tid-1) + \textbf{\em gap})$
\State $A1[tid] \gets \textrm{\textbf{max}}(A1[tid],~tmp)$
\If{$A1[tid] < best - X$}
\State $A1[tid] \gets -\infty$
\EndIf
\State $tid \gets tid+numScheduledThreads$
\EndWhile
\EndProcedure
\end{algorithmic}
\end{algorithm}

Once each thread is assigned to a cell of the anti-diagonal, our algorithm leverages in-warp thread communication to compare the values inside cells and perform the reduction, where a warp is a set of 32 threads executing the same code and sharing the same data.
All threads within a warp communicate using registers, which maximizes communication speed.
Algorithm~\ref{alg:antidiag_al} illustrates the parallel computation of the anti-diagonal.
Finally, the size of the next anti-diagonal to be computed is updated by checking if there are cells marked with $-\infty$ at the end or the start of the current anti-diagonal.
LOGAN continues computing the alignment matrix until either it reaches the end of the shortest read or the size of the current anti-diagonal is set to zero, meaning that it has satisfied the  condition. 

\begin{figure}[t]
\centering

\includegraphics[width=.5\columnwidth]{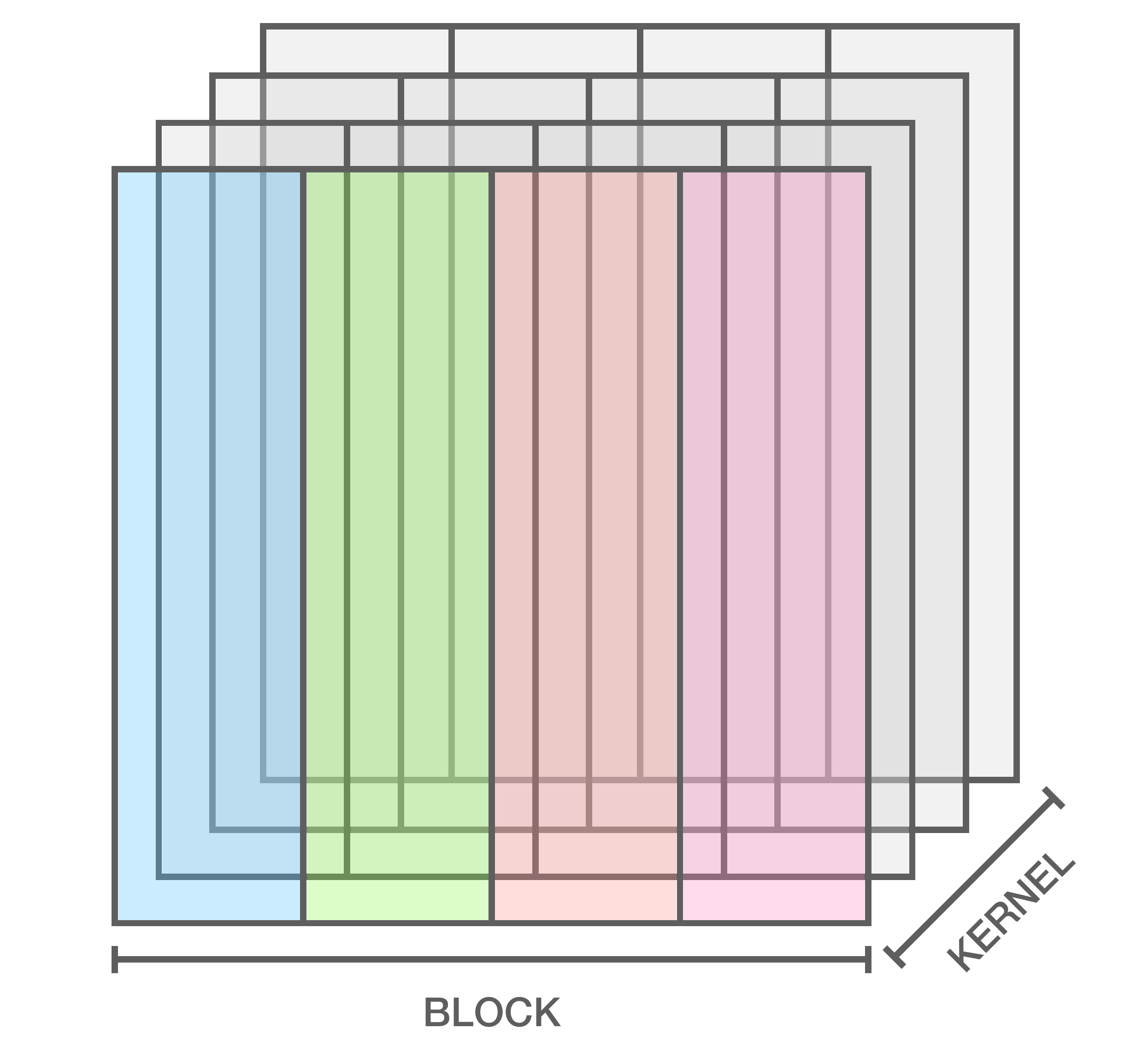}
\caption{
    Each alignment is assigned to one block. The kernel executes all the block in parallel, in order to leverage inter-alignment parallelization.
}
\label{fig:segmentation}
\end{figure}

\subsection{Inter-Sequence Parallelism}\label{sec:inter}

Intra-sequence parallelism optimizes the alignment computation for a single pair of sequences, however, it does not effectively leverage the large volume of GPU computational resources. 
We therefore exploit the GPU computational potential by designing LOGAN to align multiple pairs of sequences in parallel by 
assigning each alignment to a GPU block ---  thus taking advantage of inter-sequences parallelism. 
LOGAN schedules the number of GPU blocks based on the number of alignments needed to be performed (\Cref{fig:segmentation}).
Each NVIDIA V100s GPU block can store up to 64KB in shared-memory and performs an independent alignment.
Since only three anti-diagonals need to be stored, we could ostensibly store them into GPU shared-memory (the fastest memory available after the registers). 
However, despite the potential of reserving 64KB of memory per block, this cannot be implemented in practice as the device has only 96KB of memory per streaming multiprocessor (SM).
Given that each SM on the device can execute up to $32$ blocks in parallel, if a single block has reserved too much memory, a  SM  is forced to exchange data with the DRAM of the GPU every time it computes a block.
Furthermore, given that only a single block could fit on a single SM, the execution would be limited to a single block per SM.
Given our goal of achieving the best possible board-level utilization, we need to compute multiple blocks per SM in parallel.
Consequently, to overcome these limitations and avoid shared memory contention, LOGAN stores the three anti-diagonals of each alignment on the \ac{HBM} of the GPU.
Doing so,  removes the limitation on the number of blocks per SM, and achieves significantly more effective parallelism. 

\begin{figure}[t]
\centering
\includegraphics[width=\columnwidth]{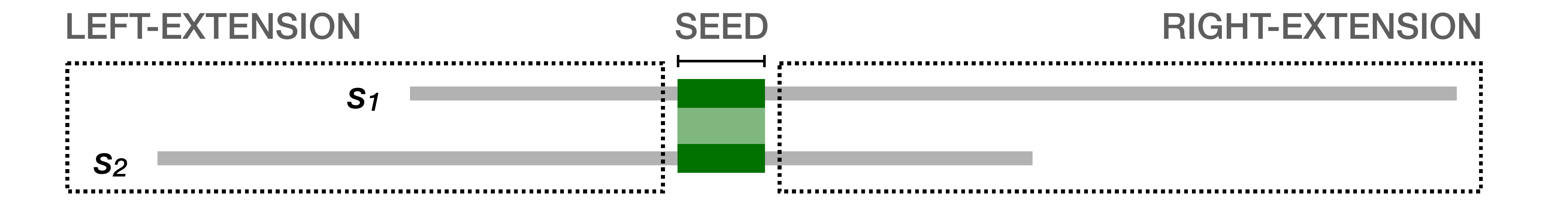}
\caption{
    In the seed-and-extend alignment paradigm, the seed location determines where the read pair is split into two different alignments: left- and right-extension.
}
\label{fig:extension}
\end{figure}

To further improve our resource utilization, LOGAN schedules a number of threads per block lower than the maximum of $1024$.
In fact, if the number of threads exceeds the anti-diagonal length, many of the threads will stall, decreasing overall performance.
Additionally, we need to store the intermediate results of the parallel reduction in shared memory to enable in-warp thread communication when computing the anti-diagonal maximum score.
Since the number of intermediate results is equal to the number of scheduled threads, reducing the number of scheduled threads per block also reduces the risk of shared memory contention.
Given that the length of each anti-diagonal is proportional to the value of $X$, our implementation schedules a number of threads proportional to $X$, significantly reducing the number of stalled threads.
This scheduling increases our performance and improves our resource utilization.
\Cref{table:parallel} shows the impact of the various degrees of parallelization implemented in the LOGAN kernel.
The first two rows of the table show the impact of intra-level parallelism over a single-thread execution, while the second two show the impact over inter-level parallelism for $100$K alignments of read pairs.
Note that intra-level parallelism improves the performance by a factor of about $9\times$, while the introduction of inter-level parallelism improves performance by an impressive factor of $22,000\times$ with respect to intra-parallelism alone.
The intra-sequence parallelism has insufficient work to consume available GPU resources, hence its impact on performance is limited compared to inter-sequence parallelism. 
Notably, inter- and intra-sequence parallelisms are complementary to each other. 
LOGAN therefore implements both to better exploit the resources of the GPU and maximize our kernel performance.
\begin{table}[t]
    \centering
    \caption{$X$-drop execution times on GPU using $X=100$ and exploiting different levels of parallelism.}
    \label{table:parallel}
    \resizebox{\columnwidth}{!}{%
	\begin{tabular}{lrrrrrr}
	\toprule
	 \multicolumn{1}{c}{\textbf{Parallelism}} &
    \multicolumn{1}{c}{\textbf{Pairs}}  & \multicolumn{1}{c}{\textbf{Threads}} & \multicolumn{1}{c}{\textbf{Blocks}}  & \multicolumn{1}{c}{\textbf{Time}} &
    \multicolumn{1}{c}{\textbf{Speed-Up}}\\
    \midrule
    None & 1 & 1 & 1 & 1.50s & - \\
    \midrule
    Intra-sequence & 1 & 128 & 1 & 0.16s & 9.3$\times$ \\
    \toprule
    Intra-sequence & 100K & 128 & 1 & 45h & - \\
    \midrule
    Intra- and inter-sequence & 100K & 128 & 100K & 7.35s & 22000.0$\times$ \\
    \bottomrule
    \end{tabular}%
    }
\end{table}
\begin{figure}[t]
    \centering
    \includegraphics[width=\columnwidth]{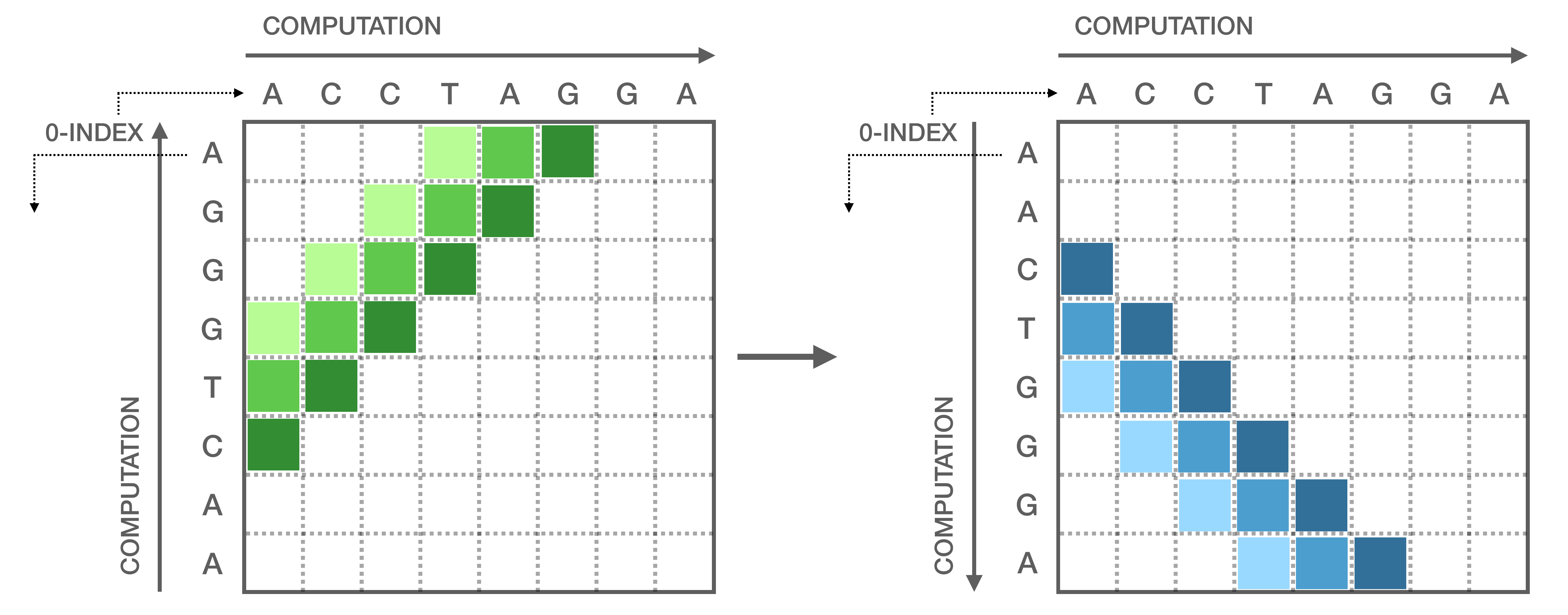}
    \caption{The query (vertical) sequence is reversed to exploit coalesced memory access on the GPU.}
    \label{fig:reversed}
\end{figure}

\begin{figure}[t]
\centering
\vspace{.2cm}
\includegraphics[width=\columnwidth]{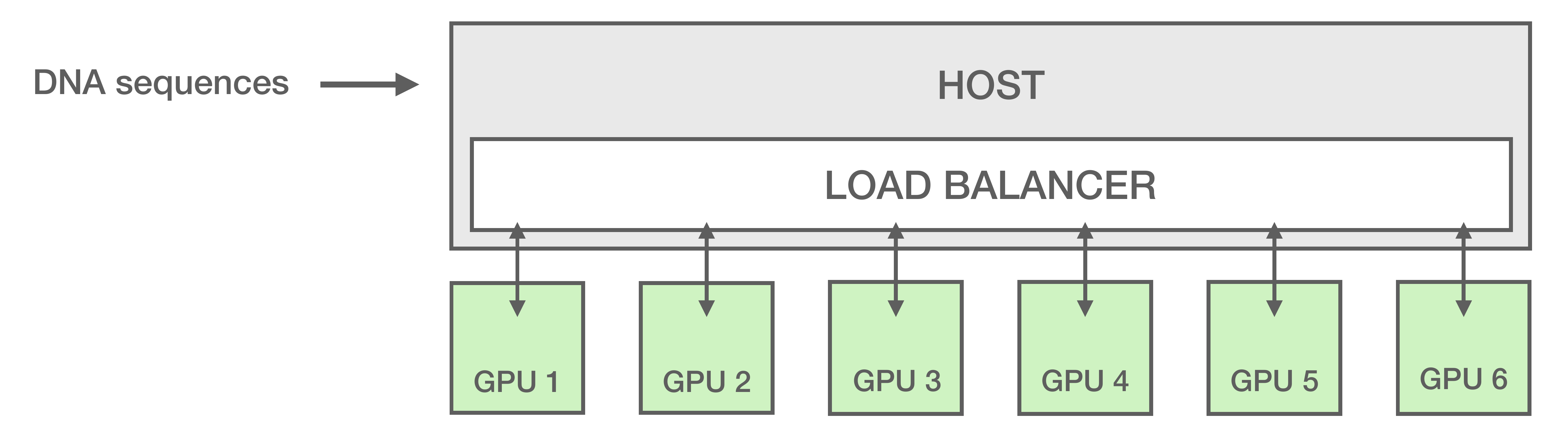}
\caption{
    Load balancing scheme of our multi-GPU implementation to distribute alignments across GPUs.
}
\label{fig:balancer}
\end{figure}

To make the LOGAN processing more efficient on the GPU, we additionally introduce CPU host optimization.
First, LOGAN loads the length of the sequences and the seed locations for each pair of sequences and  stores them into two buffers.
Each pair of sequences is split in two based on the seed's location, resulting in a {\it left-extension} pair and a {\it right-extension} pair, as shown in Figure~\ref{fig:extension}. 
Left-extension and right-extension pairs are stored into two different vectors, and these alignments are computed independently by scheduling two different streams on the GPU. 
Traditionally, when aligning two sequences, one of  them is accessed backward, resulting in memory performance degradation since  characters are read in the opposite direction of the memory.
To ensure coalesced data access on the GPU and exploit memory burst, one of the two sequences (for each given pair) is reversed, as shown in Figure~\ref{fig:reversed}.
This optimization linearizes the GPU memory data access, thus increasing performance while preserving the correct solution. 
Finally, note that kernel execution is scheduled asynchronously from the host, enabling the retrieval of alignment results as soon as they are available, instead of waiting for all the alignments to complete.


\subsection{Implementation with Multiple GPU Devices}\label{sec:multi}

To effectively exploit the available multiple GPU resources, LOGAN leverages a load balancer as shown in Figure~\ref{fig:balancer}.
This optimization allows LOGAN to run on varying GPU configurations, since it can adapt the load to the specific number of GPUs present within a given system.

The host application balances the computation by scheduling the number of alignments for each GPU.
The host switches context multiple times and then replicates the operations for each GPU to simplify its task. 
The pre-processing of the sequences occurs as in the single device implementation and the load balancer divides the sequences into different groups that are then assigned to the GPUs.
The HBM memory of the GPU represents a limiting resource for LOGAN, since in the single GPU implementation it is fully utilized.
To ensure balance, we schedule the number of alignments per GPU considering both the number of available GPUs and the length of the sequences.
Once the division is completed, the host allocates the necessary memory on the different GPUs, enabling each GPU to execute its set of alignments independently.
The host then schedules each GPU kernel to be executed in parallel and collects alignment results asynchronously.
Once the GPU devices completed their execution, the load balancer collects and organizes the results.
\section{BELLA Integration}\label{abs:ki}
To demonstrate the impact of the work, we integrate LOGAN into a long-read analysis software, called BELLA \cite{guidi2018bella}.
BELLA is a recently released, publicly-available software for long-read many-to-many overlap detection and alignment. Detecting overlaps is a crucial and computationally intense step in many long-read applications, such as {\it de novo} genome assembly and error correction. 
BELLA uses a seed-based approach for overlap detection implemented as an efficient sparse matrix-matrix multiplication (SpGEMM) kernel.
Before performing overlapping, BELLA provides a new algorithm for pruning the $k$-mers, substrings of fixed length $k$ used as seeds. The  $k$-mers are pruned because unlikely to be useful in overlap detection and their retention would cause unnecessary computational overhead and potential errors.
Once overlaps are identified, a seed-end-extend pairwise alignment step is performed to filter out spurious overlaps.
BELLA chooses the optimal $k$-mer to begin alignment extension, as illustrated in Figure~\ref{fig:extension}, through a {\it binning mechanism}, where $k$-mer locations are used to estimate the overlap length and to ``bin'' $k$-mers to form a consensus.
BELLA additionally uses the novel approach of 
separating true alignments from false positives using an {\it adaptive threshold} based on a combination of alignment techniques and probabilistic modeling.

BELLA  relies on SeqAn's $X$-drop implementation \cite{doring2008seqan} for pairwise alignment, which constitutes about $90\%$ of the overall runtime when using real data sets.
Once overlaps are computed via SpGEMM, BELLA performs the pairwise alignment and determines if the aligned pair should be kept.
The current implementation is efficient for SeqAn as the processor computes independent pairwise alignments in parallel using OpenMP~\cite{dagum1998openmp}.
However, this approach is inefficient for the GPU architecture, since it limits the amount of parallelism between alignments.
To better exploit inter-alignment parallelism, we modify BELLA to batch the entire set of alignments together and send them to the GPU devices.
The host CPU then retrieves and post-processes the alignment results. 
Our optimized BELLA version with LOGAN integration produces equivalent results as the original version.

\section{Discussion}
\label{abs:res}
\begin{figure}[t]
    \centering
    \includegraphics[width=\columnwidth]{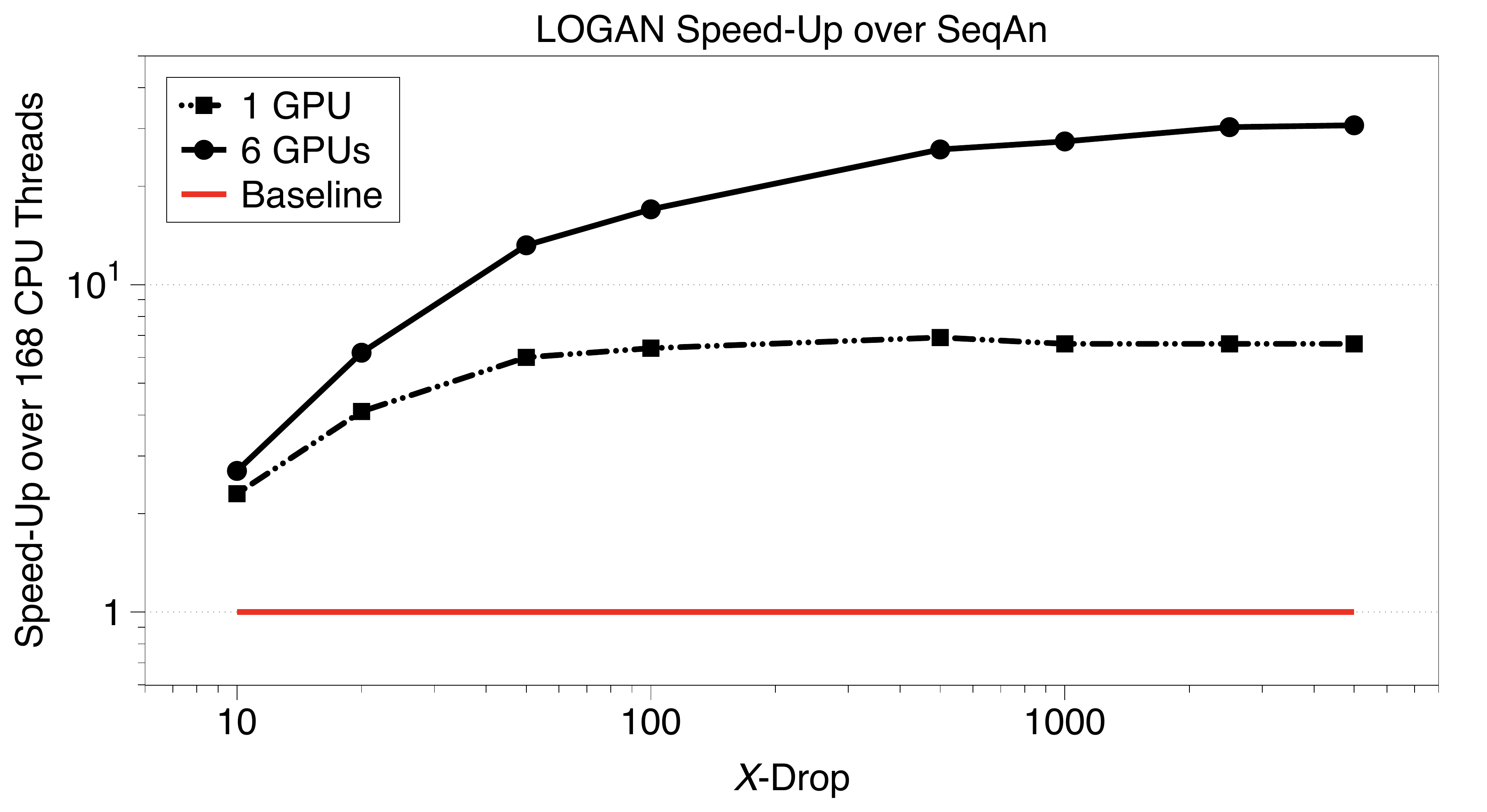}
    \caption{LOGAN's speed-up over SeqAn for 100K alignments (log-log scale). POWER9 Platform with 6 NVIDIA Tesla V100s.}
    \label{fig:performance_comparison_1}
\end{figure}
\begin{table}[t]
	\centering
	\caption{\textbf{LOGAN} and \textbf{SeqAn} execution times in seconds for 100K alignments (POWER9 platform with 6 NVIDIA Tesla V100s).}
	\label{table:performance_comparison_1}
	\begin{tabular}{rrrr}
	\toprule
		\multirow{2}{*}{\textbf{$X$-Drop}} &\textbf{SeqAn}& \textbf{LOGAN} & \textbf{LOGAN} \\
		 &\textbf{168 CPU Threads}& \textbf{1 GPU} & \textbf{6 GPU} \\
	\midrule
	    10  & 5.1  &2.2 &1.9 \\
	\midrule
        20  & 12.7 &3.1 &2.1\\
    \midrule
        50 & 29.6  &5.0 &2.2 \\
    \midrule
        100 & 45.7 &7.2 &2.7\\
    \midrule
        500& 102.6 &14.9&4.0\\
    \midrule
        1000& 133.3 &20.2&4.9\\
    \midrule
        2500& 168.0&25.3&5.6 \\
    \midrule
        5000 & 176.6 & 26.7 &5.8 \\
	\bottomrule
	\end{tabular}%
\end{table}

\begin{figure}[t]
    \centering
    \includegraphics[width=\columnwidth]{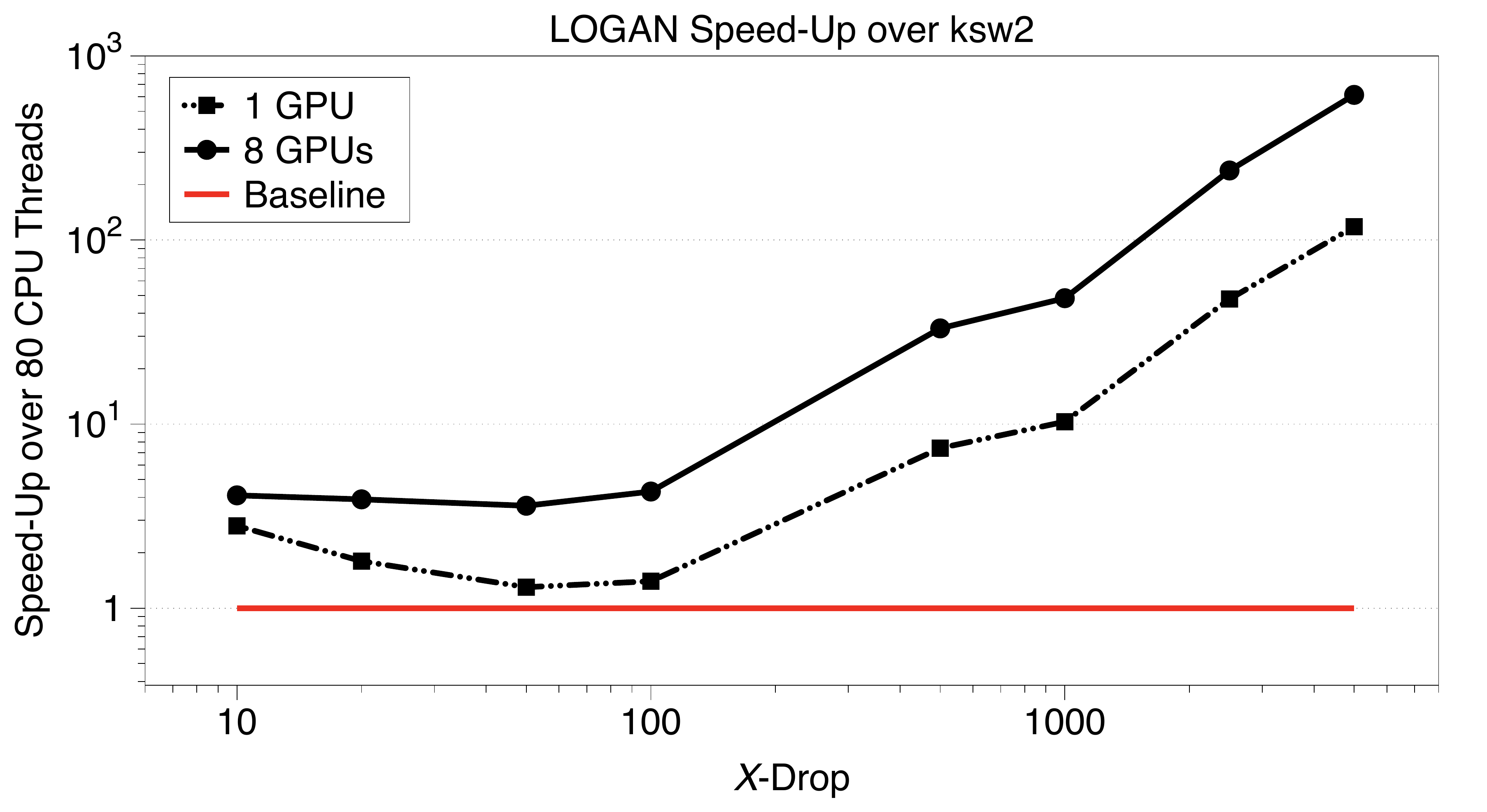}
    \caption{LOGAN's speed-up over ksw2 for 100K alignments (log-log scale). ``Skylake'' Platform with 8 NVIDIA Tesla V100s.}
    \label{fig:performance_comparison_4}
\end{figure}
\begin{table}[t]
	\centering
	\caption{\textbf{LOGAN} and \textbf{ksw2} execution times in seconds for 100K alignments (``Skylake'' platform with 8 NVIDIA Tesla V100s).}
	\label{table:performance_comparison_4}
	\begin{tabular}{rrrr}
	\toprule
		\multirow{2}{*}{\textbf{$X$-Drop}} &\textbf{ksw2}& \textbf{LOGAN} & \textbf{LOGAN} \\
		 &\textbf{80 CPU Threads}& \textbf{1 GPU} & \textbf{8 GPU }\\
	\midrule
	    10  & 6.9 & 2.5 & 1.7 \\
	\midrule
        20  & 7.0 & 3.8 & 1.8 \\
    \midrule
        50 & 7.7 & 5.8 & 2.1 \\
    \midrule
        100 & 10.4 & 7.3 & 2.4 \\
    \midrule
        500 & 113.0 & 15.2 & 3.4 \\
    \midrule
        1,000& 209.5 & 20.4 & 4.3 \\
    \midrule
        2,500& 1235.8 & 25.9 & 5.2 \\
    \midrule
        5,000& 3213.1 & 27.2 & 5.2 \\
	\bottomrule
	\end{tabular}%
\end{table}

\begin{figure}[t]
    \centering
    \includegraphics[width=\columnwidth]{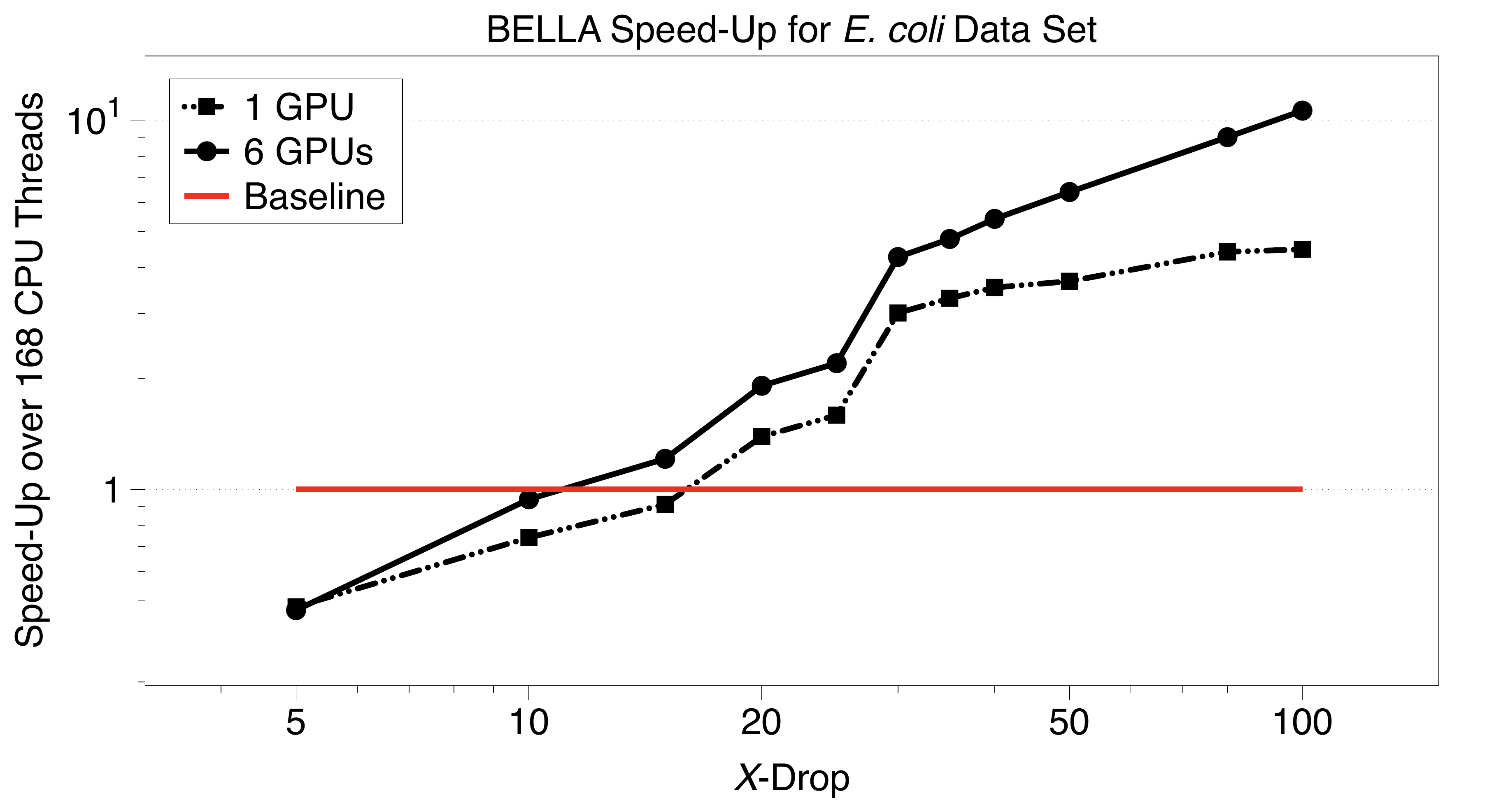}
    \caption{
    BELLA's speed-up replacing its pairwise alignment kernel (SeqAn) with LOGAN for the {\it E. coli} data set for 1.8M alignments (log-log scale). POWER9 Platform with 6 NVIDIA Tesla V100s.}
    \label{fig:performance_comparison_3}
\end{figure}
\begin{table}[t]
	\centering
	\caption{Execution times on POWER9 platform with 6 NVIDIA Tesla V100s in seconds for 1.82M alignments ({\em E. coli}).}
	\label{table:performance_comparison_3}
	\begin{tabular}{rrrr}
	\toprule
		\multirow{2}{*}{\textbf{$X$-Drop}} &\textbf{BELLA}& \textbf{LOGAN} & \textbf{LOGAN}\\ 
		 &\textbf{168 CPU Threads}& \textbf{1 GPU} & \textbf{6 GPU} \\
	\midrule
	    5 &  53.2 & 110.4 & 114.3 \\ 
	\midrule
        10 &108.6 & 146.4 & 115.3 \\ 
    \midrule
        15 &139.0 & 152.9 & 114.8 \\ 
    \midrule
        20 &226.7 & 162.7 & 118.4 \\ 
    \midrule
        25 &275.3 & 173.5 & 125.3 \\ 
    \midrule
        30 &558.0 & 185.3 & 130.6 \\
    \midrule
        35 &654.1 & 198.4 & 136.8 \\ 
    \midrule
        40 &750.1 & 212.7 & 138.4 \\ 
    \midrule
        50 &913.1 & 248.5 & 141.4 \\
    \midrule
        80 &1303.7 & 295.8 & 142.4 \\
    \midrule
        100 & 1507.1 & 336.3 & 144.5 \\
	\bottomrule
	\end{tabular}%
\end{table}

In this section, we describe the experimental settings used to evaluate the LOGAN methodology and present our performance results.

\subsection{Experimental Setting}
We first compare LOGAN against the CPU-based $X$-drop algorithm as implemented in SeqAn~\cite{doring2008seqan}. 
Next, we evaluate LOGAN against two GPU-based algorithms: the current state-of-the-art implementation of full Smith-Waterman (SW), CUDASW3++~\cite{liu2013cudasw++}, and the closest heuristics to ours proposed by Feng et al., manymap~\cite{FengIcpp2019}.
Finally, we integrate LOGAN into the BELLA long-read application to demonstrate its benefit in a real-world computation.

To compare LOGAN against SeqAn's, we generate a set of 100K read pairs with read length between 2,500 and 7,500 characters and an error rate of $\approx$15$\%$ between two reads of a given pair. 
The results were collected on a dual-socket server with two 22-core IBM POWER9 processors and 6 NVIDIA Tesla V100s (16 GB HBM2) with 512 GB DDR4 of RAM. Each processor has 21 compute cores with 4 threads per core.
Also, we compare LOGAN to minimap2's \cite{Minimap2} vectorized $Z$-drop alignment algorithm, called ksw2 \cite{suzuki2018introducing}, using the same data set of 100K pairs described above. 
Note that we conducted these comparisons on a different hardware platform: a dual-socket computer with two 20-core Intel Xeon Gold 6148 CPU processors, each running at 2.40 GHz with 384 GB DDR4 $2400$ MHz memory and 8 NVIDIA Tesla V100s (16 GB HBM2) GPUs. 
A different platform was required since the POWER9 processors are not compatible with ksw2's SSE2 SIMD instructions.
On the Intel Xeon Gold platform with 8 NVIDIA Tesla V100s, we also perform the comparison between LOGAN, CUDASW++, and manymap using the same 100K pairs as above.

Additionally, we integrate LOGAN into the BELLA long-read application~\cite{guidi2018bella}, described in \Cref{abs:ki}, by evaluating the  performance difference of replacing SeqAn with LOGAN.
For this comparison, we used a real {\em E. coli} and a synthetic {\em C. elegans} data sets, requiring 1.8M and 235M alignments, respectively.
We analyzed these experiments on the same hardware platform as SeqAn evaluation.

\subsection{Results}
\begin{figure}[t]
    \centering
    \includegraphics[width=\columnwidth]{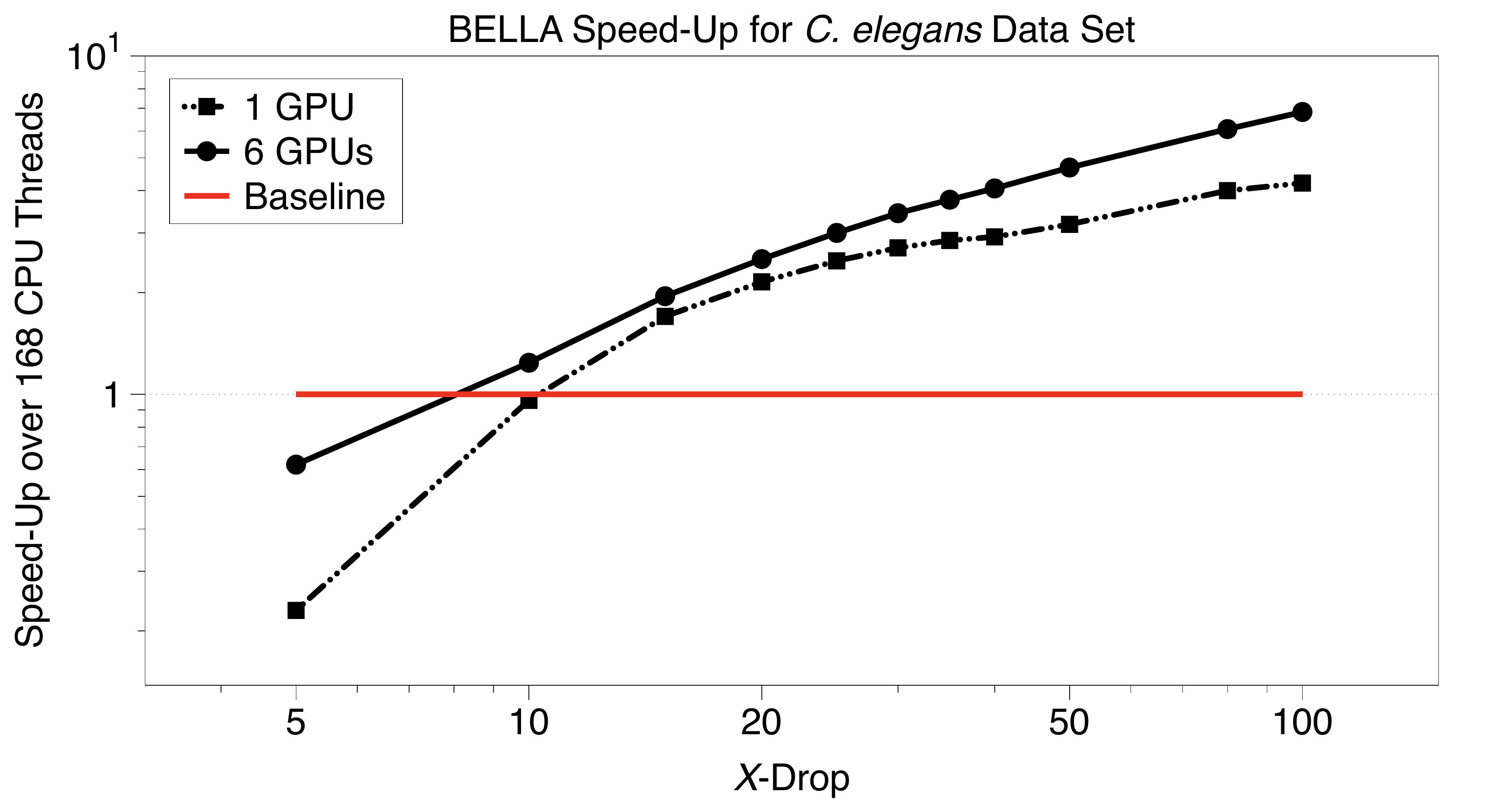}
    \caption{BELLA's speed-up replacing its pairwise alignment kernel (SeqAn) with LOGAN for the {\it C. elegans} data set for $235$M alignments (log-log scale). POWER9 Platform with 6 NVIDIA Tesla V100s.}
    \label{fig:performance_comparison_2}
\end{figure}
\begin{table}[t]
	\centering
	\caption{Execution times on POWER9 platform with 6 NVIDIA Tesla V100s in seconds for 235M alignments ({\em C. elegans}).}
	\label{table:performance_comparison_2}
	\begin{tabular}{rrrr}
	\toprule
		\multirow{2}{*}{\textbf{$X$-Drop}} &\textbf{BELLA}& \textbf{LOGAN} & \textbf{LOGAN}\\ 
		 &\textbf{168 CPU Threads}& \textbf{1 GPU} & \textbf{6 GPU} \\
	\midrule
	    5 & 131.7 & 577.1 & 213.1 \\ 
	\midrule
        10 & 723.3 & 750.2 & 579.7 \\ 
    \midrule
        15 & 1467.7 & 865.6 & 749.8 \\ 
    \midrule
        20 & 1954.8 & 908.9 & 777.0 \\ 
    \midrule
        25 & 2518.8 & 1015.5 & 838.9 \\ 
    \midrule
        30 & 3047.1 & 1125.0 & 888.0 \\
    \midrule
        35 & 3492.5 & 1226.5 & 927.0 \\ 
    \midrule
        40 & 3887.0 & 1329.0 & 955.9 \\ 
    \midrule
        50 & 4607.7 & 1449.0 & 983.7 \\
    \midrule
        80 & 6367.7 & 1593.9 & 1046.1 \\
    \midrule
        100 & 7385.3 & 1753.3 & 1080.9 \\
	\bottomrule
	\end{tabular}%
\end{table}

Figure~\ref{fig:performance_comparison_1} shows LOGAN's speed-up using both one GPU and the entire set of six GPUs compared against SeqAn's implementation using 168 threads on two POWER9 processors.  Details of the execution time are shown in Table~\ref{table:performance_comparison_1}.
Note that LOGAN's execution times remain roughly constant for large values of $X$.
In these scenarios, we can exploit the full parallelism of the GPU architecture, resulting in similar execution times.
Observe that LOGAN attains speed-ups ranging from $2.3\times$ to $6.6\times$ for a single GPU and from $2.7\times$ to $30.7\times$ using all six GPUs. 
As expected, LOGAN achieves higher speed-ups as the value of $X$ increases, since the alignment runs for a longer duration.
We also note that LOGAN multiple GPU implementation scales better for longer execution runs.
This is due to amortizing the load balancing overhead when dividing the sequences into different groups. 

Figure~\ref{fig:performance_comparison_4} presents LOGAN's performance using both 1 GPU and the entire set of 8 GPUs when compared against ksw2's CPU vectorized implementation on the Skylake processor.
Both algorithms are benchmarked using the same set of 100K alignments used to compare LOGAN and SeqAn.
Results show that LOGAN attains significant speed-ups ranging from $3.1\times$ to $120.4\times$ with a single GPU and from $3.7\times$ to $558.5\times$ using eight GPUs.
Additionally, we can observe that ksw2 performs better when aligning the sequences using a small value of $X$ and its performance degrades drastically when increasing the $X$-drop value, as shown in Table~\ref{table:performance_comparison_4}.
Given LOGAN and ksw2 implement two slightly different heuristics, we also report a  comparison based on the GCUPS metric. 
LOGAN achieves up to $181.4$ GCUPS with a single GPU for $X=5000$, while ksw2 best performance is only $77.6$ GCUPS for $X=100$.
Importantly, LOGAN always outperforms ksw2 both in terms of runtime and GCUPS, independently from their respective peak performance at different values of $X$.

\begin{figure}[t]
    \centering
    \includegraphics[width=\columnwidth]{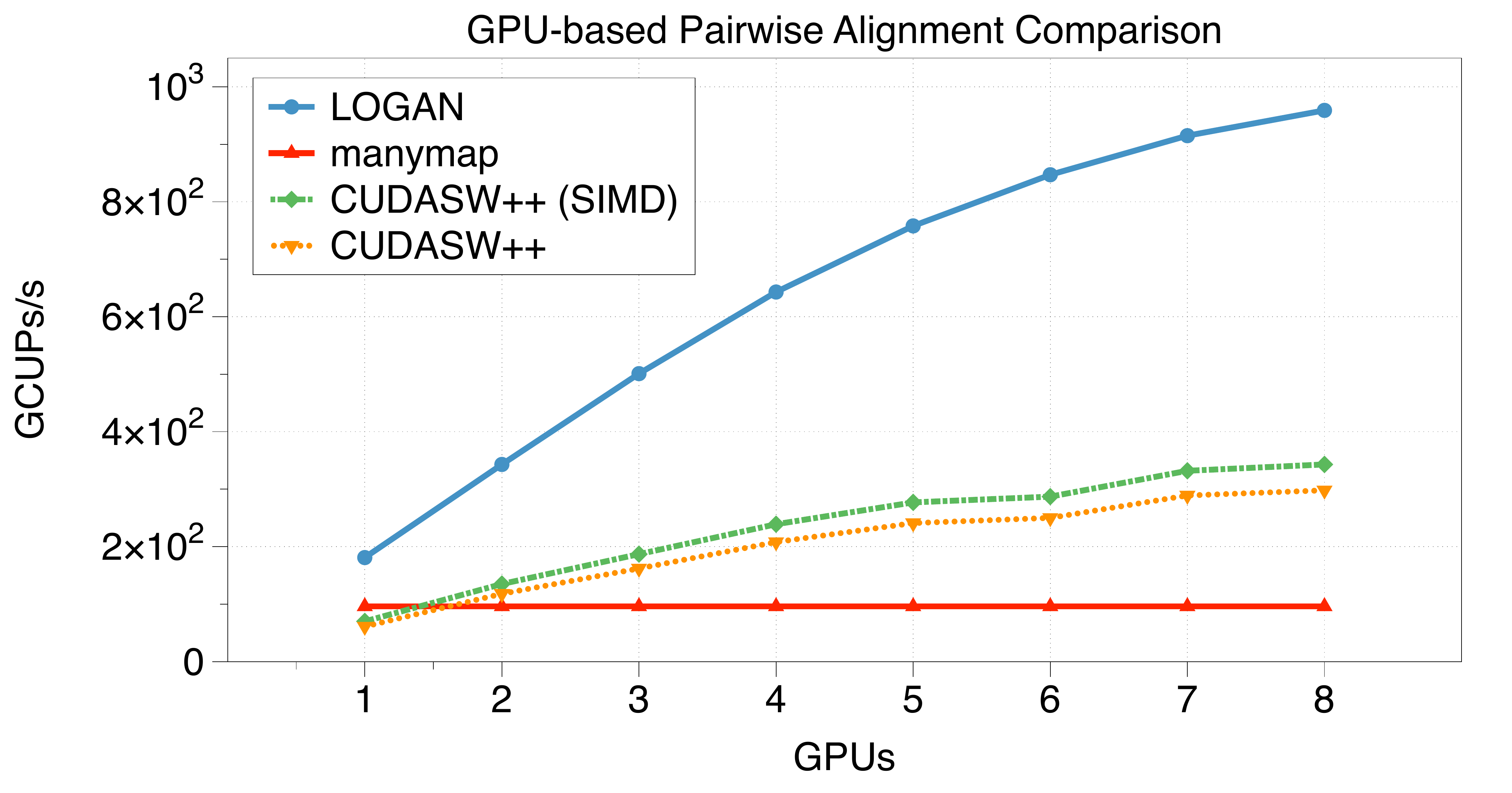}
    \caption{Comparison amongst GPU-based pairwise alignment algorithms in terms of GCUPs per second (``Skylake'' platform with 8 NVIDIA Tesla V100s). Higher is better. manymap is single GPU only, hence we report its performance as a flat line.}
    \label{fig:GPUcmp}
\end{figure}


Figure~\ref{fig:GPUcmp} illustrates LOGAN's performance compared to two GPU-based algorithms, CUDASW++ and manymap. 
Notably, each of these three implementations performs a different amount of work, therefore we report the performance in terms of GCUPS.
Furthermore, CUDASW++ uses hybrid GPU/SIMD computation by default.
We report its performance with both hybrid and GPU-only computation. 
LOGAN consistently outperforms both CUDASW++ and manymap with performance up to $181$ GCUPS on a single GPU, while CUDASW++ and manymap achieve at most $70$ and $96$ GCUPS, respectively.
Running with eight GPUs, LOGAN computes $3.2\times$ more GCUPS than GPU-only CUDASW++.


Finally, Figures~\ref{fig:performance_comparison_3} and \ref{fig:performance_comparison_2} present BELLA's performance improvements when using LOGAN as pairwise alignment kernel.
Our results show that BELLA attains significant speed-ups up to 7$\times$ and 10$\times$ on one GPU and six GPUs, respectively.
Tables~\ref{table:performance_comparison_3} and \ref{table:performance_comparison_4} show the runtime of the original software in the column named ``BELLA'' and the runtime of BELLA using LOGAN as pairwise alignment kernel in the column ``LOGAN''. 
For large values of $X$, results show that LOGAN's runtime does not drastically degrade with increasing $X$. 
Notably, BELLA operates in a context where sequences have an error rate of about $10-15\%$.
In this scenario, small values of $X$ can potentially lead to early drop-outs, even when sequences are supposed to align until the endpoints.
Up to a certain point, increasing the value of $X$ increases the number of true alignments and makes it easier to differentiate true alignments from false positives.
LOGAN's integration would allow BELLA to use larger $X$ values, resulting in higher accuracy without a notable increase in runtime.

Computing time scales linearly, however, the communication with multiple GPUs introduces an overhead that increases with the number of GPUs.



\section{LOGAN Roofline Analysis}
\label{abs:roof}
\begin{figure}[t]
\centering
\vspace{-.4cm}
\includegraphics[width=\columnwidth]{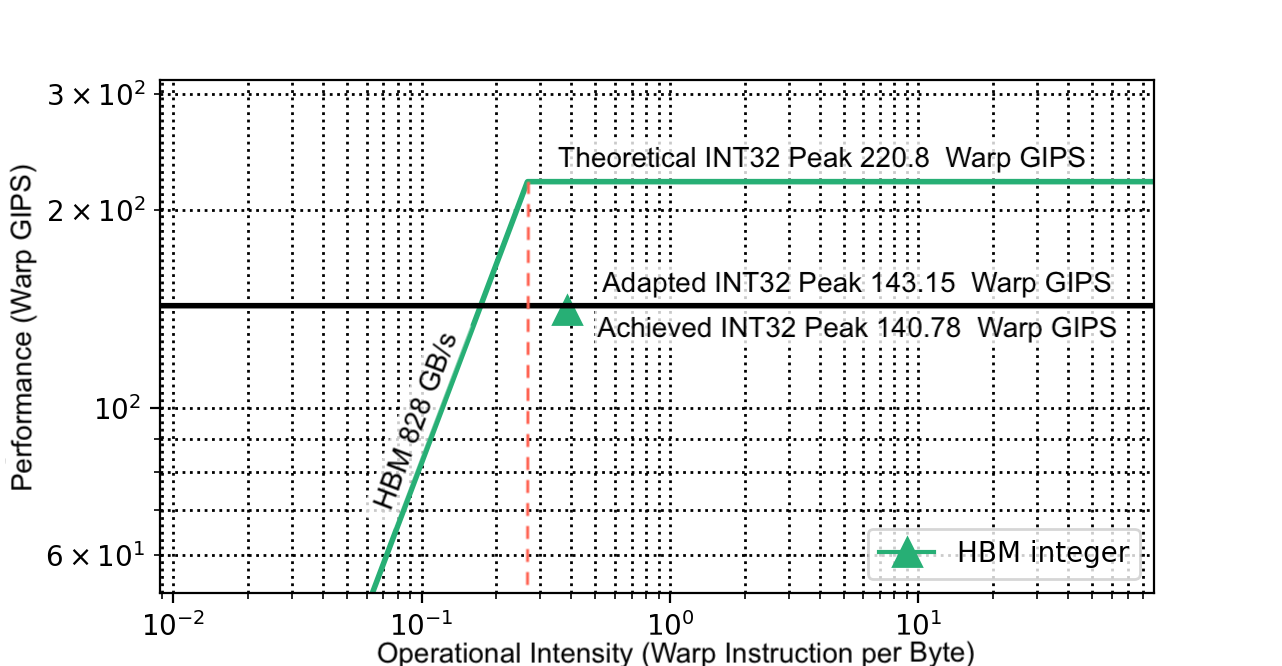}
\caption{Roofline analysis for our kernel on the NVIDIA Tesla V100 GPU performing 100K alignment and using $X=100$.}
\label{fig:roofline}
\end{figure}

In this section, we provide a detailed analysis of the optimized LOGAN GPU performance by adapting the Roofline model~\cite{CACM09_Roofline, Ding2019instructionroofline} to fit our specific computational characteristics. 
The Roofline model is a visually-intuitive method to understand the performance of a given kernel based on a {\em bound and bottleneck} analysis approach. 
The model outlines which factors affect the performance of computer systems, relating processor performance to off-chip memory traffic.
The Roofline model characterizes a kernel's performance in billions of instructions (GIPS, y-axis) as a function of its operational intensity (OI, x-axis).
We use {\em Operational Intensity} as the x-axis and, given that our kernel performs only integer operations, use billions of warp instructions per second (Warp GIPS) as the y-axis. 
{\em Operational Intensity} is defined as instructions per byte of DRAM traffic, which measures traffic between the caches and memory.
Thus, our Roofline analysis combines integer performance, operational intensity, and memory performance into a 2D log-log scale graph, as shown in Figure~\ref{fig:roofline}.

On one NVIDIA Tesla V100 GPU, 80 \ac{SM}s are available, where each \ac{SM} consists of four processing blocks, called {\em warp schedulers}.
Each warp scheduler can dispatch only one instruction per cycle.
As such, the theoretical maximum (warp-based) of instruction/s is $80\textrm{~SM} \times (4 \times \textrm{warp~scheduler}) \times (1 \times \sfrac{\textrm{instruction}}{\textrm{cycle}}) \times 1.53\textrm{~GHz} = 489.6 $ \textrm{GIPS}.
Besides, each processing block contains 16 FP32 cores, 8 FP64 cores, and 16 INT32 cores.
The maximum attainable integer performance is $\sfrac{16}{32} \times 489.6 = 220.8$ integer warp GIPS since 16 INT32 cores can only support 16 threads out of 32 threads in one warp.
Peak Performance is upper bounded by both the theoretical INT32 peak rate and the peak memory bandwidth, which define the green line in the plot.
The actual Warp \ac{GIPS} depends on the operational intensity and the ceiling line determines the limit of the actual performance. 
A kernel is memory-bound if the Warp GIPS are limited by the memory bandwidth (left of the red dotted line), and is compute-bound if limited by the hardware performance limit (right of the red dotted line).
This maximum attainable performance represents a ceiling in the Roofline model plot for the considered GPU platform and is independent from the executed algorithm.
To adapt this ceiling to the $X$-drop algorithm, we use the following formula:

\begin{equation}
\label{eqn:ceiling}
Ceiling=\frac{1}{N}\sum_{i=1}^{N}\frac{f \times N_{op,i} \times B}{\lceil (T \times B)/\mathit{MAX_{R}}\rceil}
\end{equation}

\Cref{eqn:ceiling} defines a new ceiling by averaging the number of cells that GPU can compute in parallel.
$N$ indicates the total number of parallel iterations for a given algorithm, $f$ is the theoretical ceiling (220.8 warp GIPS), $B$ is the number of scheduled blocks, $N_{op,i}$ indicates the number of operations that need to be computed at each iteration, $T$ is the number of scheduled threads per block, and, finally, $\mathit{MAX_R}$ indicates the number of INT32 cores available.
LOGAN's overall performance behavior is shown in Figure~\ref{fig:roofline}.
Result show the operational intensity of our kernel on the HBM memory of the GPU, indicating that we are not memory bound and that we are bound by the adapted theoretical ceiling.
In other words, the operational intensity of our kernel is high enough to be in the compute-bound area of the Roofline, thus it is not limited by the HBM memory.
Note that the optimized performance of our algorithm is very close to the adapted theoretical ceiling.
Considering that the adapted ceiling does not take into account memory latency, the results of our implementation are extremely close to the maximum achievable performance.
Therefore, LOGAN represents a near-optimal implementation of the $X$-drop algorithm and it is only limited by the compute capability of the GPU.

\section{Conclusions}
\label{abs:concl}
Our work presents LOGAN, the first high-performance multi-GPU implementation of the $X$-drop alignment algorithm. $X$-drop is employed in several important genomics applications, however it is particularly challenging for GPU parallelization due to its adaptive banding and continual termination checks.

Detailed results and analyses show significant performance acceleration using our novel optimization approach. LOGAN demonstrated runtime improvements of up to $30.7\times$ using six GPUs, compared with the original CPU algorithm.  
Additionally, results show speed-ups up to $614.4\times$ using six GPUs compared with the SIMD vectorized ksw2 algorithm, which implements a similar heuristics. Finally, LOGAN integration resulted in performance improvement up to $10.7\times$ on BELLA, a real-world many-to-many long-read overlapper and aligner.


Finally, our work provided an adaptation of the Roofline model that captures the unique aspects of our computation in the context of the underlying GPU hardware configuration.  Roofline analysis demonstrates that our $X$-drop design methodology results in near-optimal performance.  Our overall results show that our optimized kernel is flexible, efficient, and can be easily integrated into long-read application performing pairwise alignment.

Future work will focus on reducing LOGAN's load balancing overhead, to enable linear performance improvements with increasing GPU counts independent of the value of $X$.
Given our implementation can be easily adapted to solve other similar problems, we also plan to extend LOGAN to support protein alignment and expect the $X$-drop algorithm to be effective in protein homology searches.

\section*{Acknowledgments}
We would like to thank Francesco Peverelli and Muaaz Awan for useful suggestions and valuable discussions.

This work is supported by the Advanced Scientific Computing Research (ASCR) program within the Office of Science of the DOE under contract number DE-AC02-05CH11231. This research was also supported by the Exascale Computing Project (17-SC-20-SC), a collaborative effort of the U.S. Department of Energy Office of Science and the National Nuclear Security Administration.

We used resources of the NERSC supported by the Office of Science of the DOE under Contract No. DEAC02-05CH11231. 
This research also used resources of the Oak Ridge Leadership Computing Facility at the Oak Ridge National Laboratory, which is supported by the Office of Science of the U.S. Department of Energy under Contract No. DE-AC05-00OR22725.

\bibliography{paper}

\begin{thebibliography}{10}
\providecommand{\url}[1]{#1}
\csname url@samestyle\endcsname
\providecommand{\newblock}{\relax}
\providecommand{\bibinfo}[2]{#2}
\providecommand{\BIBentrySTDinterwordspacing}{\spaceskip=0pt\relax}
\providecommand{\BIBentryALTinterwordstretchfactor}{4}
\providecommand{\BIBentryALTinterwordspacing}{\spaceskip=\fontdimen2\font plus
\BIBentryALTinterwordstretchfactor\fontdimen3\font minus
  \fontdimen4\font\relax}
\providecommand{\BIBforeignlanguage}[2]{{%
\expandafter\ifx\csname l@#1\endcsname\relax
\typeout{** WARNING: IEEEtran.bst: No hyphenation pattern has been}%
\typeout{** loaded for the language `#1'. Using the pattern for}%
\typeout{** the default language instead.}%
\else
\language=\csname l@#1\endcsname
\fi
#2}}
\providecommand{\BIBdecl}{\relax}
\BIBdecl

\bibitem{needleman1970general}
S.~B. Needleman and C.~D. Wunsch, ``A general method applicable to the search
  for similarities in the amino acid sequence of two proteins,'' \emph{Journal
  of molecular biology}, vol.~48, no.~3, pp. 443--453, 1970.

\bibitem{smith1981identification}
T.~F. Smith, M.~S. Waterman \emph{et~al.}, ``Identification of common molecular
  subsequences,'' \emph{Journal of molecular biology}, vol. 147, no.~1, pp.
  195--197, 1981.

\bibitem{zhang2000greedy}
Z.~Zhang, S.~Schwartz, L.~Wagner, and W.~Miller, ``A greedy algorithm for
  aligning {DNA} sequences,'' \emph{Journal of Computational biology}, vol.~7,
  no. 1-2, pp. 203--214, 2000.

\bibitem{altschul1990basic}
S.~F. Altschul, W.~Gish, W.~Miller, E.~W. Myers, and D.~J. Lipman, ``Basic
  local alignment search tool,'' \emph{Journal of molecular biology}, vol. 215,
  no.~3, pp. 403--410, 1990.

\bibitem{kielbasa2011adaptive}
S.~M. Kie{\l}basa, R.~Wan, K.~Sato, P.~Horton, and M.~C. Frith, ``Adaptive
  seeds tame genomic sequence comparison,'' \emph{Genome research}, vol.~21,
  no.~3, pp. 487--493, 2011.

\bibitem{schwartz2003human}
S.~Schwartz, W.~J. Kent, A.~Smit, Z.~Zhang, R.~Baertsch, R.~C. Hardison,
  D.~Haussler, and W.~Miller, ``Human--mouse alignments with {BLASTZ},''
  \emph{Genome research}, vol.~13, no.~1, pp. 103--107, 2003.

\bibitem{li2018minimap2}
H.~Li, ``Minimap2: pairwise alignment for nucleotide sequences,''
  \emph{Bioinformatics}, vol.~34, no.~18, pp. 3094--3100, 2018.

\bibitem{frith2010parameters}
M.~C. Frith, M.~Hamada, and P.~Horton, ``Parameters for accurate genome
  alignment,'' \emph{BMC bioinformatics}, vol.~11, no.~1, p.~80, 2010.

\bibitem{guidi2018bella}
G.~Guidi, M.~Ellis, D.~Rokhsar, K.~Yelick, and A.~Bulu{\c{c}}, ``{BELLA}:
  Berkeley efficient long-read to long-read aligner and overlapper,''
  \emph{bioRxiv}, p. 464420, 2018.

\bibitem{farrar2006striped}
M.~Farrar, ``Striped {Smith--Waterman} speeds database searches six times over
  other {SIMD} implementations,'' \emph{Bioinformatics}, vol.~23, no.~2, pp.
  156--161, 2006.

\bibitem{langmead2012fast}
B.~Langmead and S.~L. Salzberg, ``Fast gapped-read alignment with {Bowtie 2},''
  \emph{Nature methods}, vol.~9, no.~4, p. 357, 2012.

\bibitem{szalkowski2008swps3}
A.~Szalkowski, C.~Ledergerber, P.~Kr{\"a}henb{\"u}hl, and C.~Dessimoz,
  ``{SWPS3}--fast multi-threaded vectorized {Smith-Waterman} for {IBM Cell/BE}
  and x86/{SSE2},'' \emph{BMC research notes}, vol.~1, no.~1, p. 107, 2008.

\bibitem{liu2013cudasw++}
Y.~Liu, A.~Wirawan, and B.~Schmidt, ``{CUDASW++} 3.0: accelerating
  {Smith-Waterman} protein database search by coupling {CPU} and {GPU SIMD}
  instructions,'' \emph{BMC bioinformatics}, vol.~14, no.~1, p. 117, 2013.

\bibitem{mr-cudasw2014}
A.~Muhammadzadeh, ``{MR-CUDASW} – {GPU} accelerated {Smith-Waterman}
  algorithm for medium-length (meta)genomic data,'' Master's thesis, University
  of Saskatchewan, Saskatchewan, July 2014.

\bibitem{li2007160}
I.~T. Li, W.~Shum, and K.~Truong, ``160-fold acceleration of the
  {Smith-Waterman} algorithm using a field programmable gate array ({FPGA}),''
  \emph{BMC bioinformatics}, vol.~8, no.~1, p. 185, 2007.

\bibitem{di2017architectural}
L.~Di~Tucci, K.~O'Brien, M.~Blott, and M.~D. Santambrogio, ``Architectural
  optimizations for high performance and energy efficient {Smith-Waterman}
  implementation on {FPGAs} using opencl,'' in \emph{Design, Automation \& Test
  in Europe Conference \& Exhibition (DATE), 2017}.\hskip 1em plus 0.5em minus
  0.4em\relax IEEE, 2017, pp. 716--721.

\bibitem{turakhia2018darwin}
Y.~Turakhia, G.~Bejerano, and W.~J. Dally, ``Darwin: A genomics co-processor
  provides up to 15,000 x acceleration on long read assembly,'' in \emph{ACM
  SIGPLAN Notices}, vol.~53, no.~2.\hskip 1em plus 0.5em minus 0.4em\relax ACM,
  2018, pp. 199--213.

\bibitem{FengIcpp2019}
Z.~Feng, S.~Qiu, L.~Wang, and Q.~Luo, ``Accelerating long read alignment on
  three processors,'' in \emph{Proceedings of the 48th International Conference
  on Parallel Processing}, ser. ICPP 2019.\hskip 1em plus 0.5em minus
  0.4em\relax New York, NY, USA: ACM, 2019, pp. 71:1--71:10.

\bibitem{Minimap2}
H.~Li, ``{Minimap2: pairwise alignment for nucleotide sequences},''
  \emph{Bioinformatics}, vol.~34, no.~18, pp. 3094--3100, 05 2018.

\bibitem{suzuki2018introducing}
H.~Suzuki and M.~Kasahara, ``Introducing difference recurrence relations for
  faster semi-global alignment of long sequences,'' \emph{BMC bioinformatics},
  vol.~19, no.~1, p.~45, 2018.

\bibitem{doring2008seqan}
A.~D{\"o}ring, D.~Weese, T.~Rausch, and K.~Reinert, ``{SeqAn} an efficient,
  generic {C++} library for sequence analysis,'' \emph{BMC bioinformatics},
  vol.~9, no.~1, p.~11, 2008.

\bibitem{dagum1998openmp}
L.~Dagum and R.~Menon, ``{OpenMP}: An industry-standard {API} for shared-memory
  programming,'' \emph{Computing in Science \& Engineering}, no.~1, pp. 46--55,
  1998.

\bibitem{CACM09_Roofline}
S.~Williams, A.~Waterman, and D.~Patterson, ``{Roofline: An Insightful Visual
  Performance Model for Multicore Architectures},'' \emph{Commun. ACM},
  vol.~52, no.~4, 2009.

\bibitem{Ding2019instructionroofline}
N.~Ding and S.~Williams, ``An instruction roofline model for gpus,'' \emph{2019
  IEEE/ACM Performance Modeling, Benchmarking and Simulation of High
  Performance Computer Systems (PMBS)}, 2019.

\end{thebibliography}
\bibliographystyle{IEEEtran}
\end{document}